\documentclass[aps,prx,twocolumn,reprint,longbibliography]{revtex4-2}

\usepackage{amsmath}
\usepackage{amssymb}
\usepackage{graphicx}
\usepackage{hyperref}
\usepackage{physics}
\usepackage[capitalize]{cleveref}

\DeclareMathOperator{\sgn}{sgn}
\newcommand*{\citeref}[1]{Ref.~\onlinecite{#1}}

\newcommand{\AppendixName}{Appendix}
\newcommand*{\crefappendix}[2]{Appendix~\ref{#1}}
\newcommand*{\crefeq}[2]{\cref{#1}}
\newcommand*{\creffig}[2]{\cref{#1}}

\begin{document}

\title{Markov Chain Monte Carlo in Tensor Network Representation}

\author{Synge Todo}
\email{wistaria@phys.s.u-tokyo.ac.jp}
\affiliation{Department of Physics,\! The University of Tokyo,\! Tokyo 113-0033,\! Japan}
\affiliation{Institute for Physics of Intelligence,\! The University of Tokyo,\! Tokyo 113-0033,\! Japan}
\affiliation{Institute for Solid State Physics,\! The University of Tokyo,\! Kashiwa,\! 277-8581,\! Japan}

\date{\today}

\begin{abstract}
  Markov chain Monte Carlo (MCMC) is a powerful tool for sampling from complex probability distributions.
  Despite its versatility, MCMC often suffers from strong autocorrelation and the negative sign problem, leading to slowing down the convergence of statistical error.
  We propose a novel MCMC formulation based on tensor network representations to reduce the population variance and mitigate these issues systematically.
  By introducing stochastic projectors into the tensor network framework and employing Markov chain sampling, our method eliminates the systematic error associated with low-rank approximation in tensor contraction while maintaining the high accuracy of the tensor network method.
  We demonstrate the effectiveness of the proposed method on the two-dimensional Ising model, achieving an exponential reduction in statistical error with increasing bond dimension cutoff.
  Furthermore, we address the sign problem in systems with negative weights, showing significant improvements in average signs as bond dimension cutoff increases.
  The proposed framework provides a robust solution for accurate statistical estimation in complex systems, paving the way for broader applications in computational physics and beyond.
\end{abstract}

\maketitle


Markov chain Monte Carlo (MCMC) is a flexible algorithm for generating random samples according to any arbitrary multidimensional probability distribution~\cite{newmanMonteCarloMethods1999,landauGuideMonteCarlo2014}.
It has been widely applied to simulations of many-body systems in statistical physics, particle physics, and biophysics, as well as for statistical estimation, Bayesian inference, optimization problems, etc.
MCMC generates random samples using a stochastic process called a Markov chain.
The Markov chain must satisfy the global balance according to the target distribution and the ergodicity in the phase space.
Although an MCMC method that satisfies these two conditions guarantees unbiased results in infinite time in principle, the MCMC dynamics in the phase space must be fast for the technique to work in practice~\cite{sokalMonteCarloMethods1997,robertMonteCarloStatistical2004,suwa2024}.
The variance of an averaged quantity, \(\sigma^2\), along a Markov chain is generally given by
\begin{equation}
  \sigma^2 \approx \frac{2 \sigma_0^2 \tau_{\text{int}}}{M},
  \label{eq:variance}
\end{equation}
where \(\sigma_0^2\) is the population variance of the raw time-series data, \(\tau_{\text{int}}\) is the autocorrelation time, and \(M\) is the number of Monte Carlo (MC) steps.

The MCMC method has an intrinsic difficulty that is even more severe than autocorrelation, called the negative sign problem.
For example, the Boltzmann weights sometimes become negative, or even complex, in the path-integral representation for frustrated quantum magnets, and they can not be considered (unnormalized) probabilities for the Monte Carlo sampling anymore; one has to simulate an alternative system, whose weight function is the absolute value of the original one, and take into account the sign effect via the reweighting technique~\cite{lohSignProblemNumerical1990,ferrenberg1988, mungerReweightingMonteCarlo1991}.
In many interesting systems, such as frustrated quantum magnets, itinerant electron systems, and real-time unitary evolution, however, the average sign, \(S\), the average of the reweighting factor, becomes exponentially small for larger system sizes, lower temperatures, or longer real-time integration.
As a result, the r.h.s. in \cref{eq:variance} gets an extra factor \(S^{-2}\), which ultimately leads to an explosive growth of the statistical error.

The increase of variance can be interpreted differently; due to the correlations, the number of samples that can be considered statistically independent decreases from \(M\) to \(M S^2/2\tau_{\text{int}}\).
So far, many efforts have been made to improve the statistics in MCMC sampling: non-local block updates~\cite{swendsenNonuniversalCriticalDynamics1987, duaneHybridMonteCarlo1987, evertzClusterAlgorithmVertex1993, boninsegniWormAlgorithmContinuousSpace2006, friasperezCollectiveMonteCarlo2023, chenTensorNetworkMonte2025}, extended ensemble methods~\cite{jankeMulticanonicalMonteCarlo1998, hukushimaExchangeMonteCarlo1996, diaconisAnalysisNonreversibleMarkov2000, turitsynIrreversibleMonteCarlo2011}, irreversible transition kernels~\cite{creutzOverrelaxationMonteCarlo1987, suwaMarkovChainMonte2010, bernardEventchainMonteCarlo2009, michelGeneralizedEventchainMonte2014}, mitigation of negative sign problem~\cite{hangleiterEasingMonteCarlo2020, klassenHardnessEaseCuring2020, levyMitigatingSignProblem2021, MurotaT2025}, etc.
However, almost all previous attempts have focused on recovering the effective sample size.

In this paper, we propose a novel approach for improving the efficiency of MCMC from a different perspective; we consider systematically reducing the population variance of physical quantities.
One might think that the population variance is not affected by the details of the sampling scheme since it is determined by the target distribution, e.g., the canonical ensemble, which reflects the system's physical characteristics.
However, this intuition is not that correct.
The population variance can change depending on the representation of the partition function---the definition of {\it configurations} and associated weights on which MCMC sampling is performed.
One example is the so-called improved estimator in the cluster and worm algorithms, in which the partition function and the estimators of physical quantities are defined in terms of graphs, and, in some cases, the population variance is reduced drastically~\cite{wolffCollectiveMonteCarlo1989, wolffCriticalSlowing1990, hasenbuschImprovedEstimatorsCluster1990, niedermayerImprovingImprovedEstimator1990, bakerRenormalizedCouplingConstant1995, horitaUpperLowerCritical2017, suwaLiftedDirectedwormAlgorithm2022}.
Unfortunately, there has been no systematic research in this direction so far.

We propose an MCMC formulation in a tensor network representation in the following, which systematically improves the population variance.
Although our method relies on the low-rank approximation in the conversion to a tensor network representation, the systematic error due to that approximation is eliminated by MCMC sampling.
By adopting the tensor network representation, we demonstrate that statistical error decreases exponentially by increasing bond dimension cutoff in the low-rank approximation and that even the negative sign problem is prevented with polynomial computational time.


Recently, tensor network methods have attracted attention as a promising numerical technique.
Tensor networks, a versatile data structure for compactly representing large matrices or tensors, have applications in various fields~\cite{schollwockDensitymatrixRenormalizationGroup2011, orusPracticalIntroductionTensor2014, xiangDensityMatrixTensor2023}.
Beyond their first applications in statistical mechanics for many-body systems~\cite{whiteDensityMatrixFormulation1992,nishinoCornerTransferMatrix1996, levinTensorRenormalizationGroup2007}, they are employed for diverse tasks such as compressing machine learning models~\cite{stoudenmireSupervisedLearningQuantumInspired2017, hanUnsupervisedGenerativeModeling2018, gaoCompressingDeepNeural2020}, simulating complex quantum circuits~\cite{liuClosingQuantumSupremacy2021, panSimulatingSycamoreQuantum2021, seitzSimulatingQuantumCircuits2023, ayralDensityMatrixRenormalizationGroup2023}, and providing high-precision approximate solutions for partial differential equations~\cite{yeQuantuminspiredMethodSolving2022, gourianovQuantuminspiredApproachExploit2022, kornevNumericalSolutionIncompressible2023, sakuraiLearningParameterDependence2025}.

There are two main approaches in the application of tensor networks.
One is the variational (or Hamiltonian) approach, where tensor networks are used as approximate functions with high representability.
They are used as variational wave functions for quantum many-body systems or as architectures for generative models in machine learning.
The other is called the renormalization (or Lagrangian) approach.
For example, the partition function of various classical lattice models and quantum circuits can be expressed precisely as tensor networks.
Then, the partition function or quantum amplitude can be obtained as the contraction of the whole network.

However, efficient tensor network contraction is a non-trivial task in either approach~\cite{schutskiSimpleHeuristicsEfficient2020, grayHyperoptimizedCompressedContraction2024}.
Generally, exact contraction requires exponential computational cost for high-dimensional tensor networks.
Therefore, a low-rank approximation based on the singular value decomposition (SVD) is usually employed during the contraction.
Since the introduction of the renormalization group methods such as density-matrix renormalization group~\cite{whiteDensityMatrixFormulation1992}, corner transfer matrix renormalization group~\cite{nishinoCornerTransferMatrix1996}, and tensor renormalization group method~\cite{levinTensorRenormalizationGroup2007}, various improvements have been proposed, such as generalization to higher-dimensional systems~\cite{xieCoarsegrainingRenormalizationHigherorder2012,adachiAnisotropicTensorRenormalization2020, panContractingArbitraryTensor2020} and fermionic systems~\cite{guGrassmannTensorNetwork2010, guEfficientSimulationGrassmann2013, akiyamaMoreGrassmannTensor2021, akiyamaTensorRenormalizationGroup2024}, improvements in computational cost~\cite{adachiAnisotropicTensorRenormalization2020, panContractingArbitraryTensor2020}, incorporation of environmental effects~\cite{orusSimulationTwodimensionalQuantum2009,xieSecondRenormalizationTensorNetwork2009,adachiBondweightedTensorRenormalization2022}, and removal of local correlations~\cite{evenblyAlgorithmsTensorNetwork2017, yangLoopOptimizationTensor2017, hauruRenormalizationTensorNetworks2018}.
The bond dimension cutoff, \(d\), which denotes the number of singular values retained in the low-rank approximation, is crucial in controlling the systematic error in the approximate contraction of tensor networks.
The sum of squares of discarded singular values gives the squared Frobenius norm of the difference between the original and the approximated tensors.
The systematic error decreases rapidly as the bond dimension cutoff increases, but the computational cost increases as some power of \(d\).
In addition, since multiple approximations are combined during the renormalization, a decrease in systematic error is often not monotonic, making the extrapolation to the \(d \rightarrow \infty\) limit difficult.

\begin{figure}[tbp]
  \centering
  \includegraphics[width=0.7\linewidth]{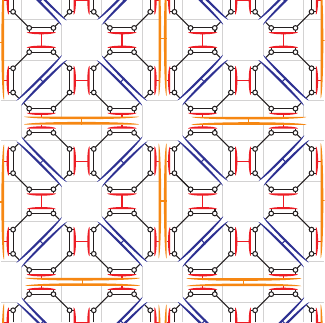}
  \caption{%
    Tensor network of the Levin-Nave TRG in the projector formulation.
    Black circles represent the tensors after the initial (exact) SVD.
    Red, blue, and orange pairs of three-leg tensors represent the projectors for the successive low-rank decompositions.
    See \crefappendix{sec:trg}{S3} for more details.
  }\label{fig:trgp-8}
\end{figure}

Let's consider the tensor renormalization group (TRG) method proposed by Levin and Nave in 2007~\cite{levinTensorRenormalizationGroup2007} as an example.
In TRG, at each renormalization step, a four-leg tensor is decomposed into two three-leg tensors, and then four three-leg tensors are contracted to form a renormalized four-leg tensor.
By repeating this procedure, all tensors are contracted to produce the partition function (see \creffig{fig:trg}{S3} in \crefappendix{sec:trg}{S3}).

The successive low-rank decompositions in the renormalization process can be replaced by the insertion of projectors, as shown in \cref{fig:trgp-8}, where each step is described by
\begin{align}
 A^* B \approx A^* P B,
 \label{eq:low-rank-decomposition}
\end{align}
where \(P = W_R W_L^* = \sum_{i=1}^d \eta_i \xi_i^* \in \mathbb{C}^{r \times r}\) is a projector to \(d\) dimensions, \(A\) and \(B\) are the tensors to be approximated, and \(r\) denotes the bond dimension between \(A^*\) and \(B\).
By choosing largest \(d\) singular values in SVD of \(A^* B = U \Sigma V^* = \sum_{i=1}^r c_i \mu_i \nu_i^*\), the projector corresponding to the best rank-\(d\) approximation is defined by \(\xi_i = c_i^{-1/2} A \mu_i\) and \(\eta_i= c_i^{-1/2} B \nu_i\) \footnote{Here, we consider the full-rank case, that is, we assume that the rank of \(A^* B\) is the same as the bond dimension \(r\) between \(A^*\) and \(B\). See \crefappendix{sec:svd}{S2} for rank-deficient cases.} (see \crefappendix{sec:svd}{S2} for more details).
The total number of projectors, denoted by \(N_\text{p}\) hereafter, is the number of repetitions of low-rank approximations and is in the same order as the number of initial tensors.
Note that a pair of \(W_R\) and \(W_L\) that make up a single projector may be far apart in the final tensor network, as the projectors are inserted recursively (\cref{fig:trgp-8}).
Other SVD-based algorithms, such as TEBD~\cite{vidalEfficientSimulationOneDimensional2004, verstraeteMatrixProductDensity2004}, ATRG~\cite{adachiAnisotropicTensorRenormalization2020}, BTRG~\cite{adachiBondweightedTensorRenormalization2022}, and CATN~\cite{panContractingArbitraryTensor2020}, can be rewritten similarly in the projector formulation.


The core technique in our proposal is to replace the projectors with stochastic ones and sample them by utilizing MCMC.
In \citeref{ferrisUnbiasedMonteCarlo2015}, Ferris proposed a seminal sampling scheme that selects \(d\) rank-1 projectors among \(r\) according to a certain weight instead of choosing the optimal rank-\(d\) projector in the process of the tensor renormalization group, where the number of possible combinations is given by \(n_\text{c}=\binom{r}{d}\).
The sampling weights are set so that the rank-1 projectors corresponding to the larger singular values are selected with higher probabilities.
A scale factor is introduced for each rank-1 projector so that the random average of stochastic projectors becomes the identity operator~\footnote{Although the inserted tensor is not exactly a projector due to the scale factors, it is referred to as a ``projector'' in the present paper for simplicity.}:
\begin{align}
  \langle W_R(\theta) W_L^*(\theta) \rangle_{\theta} = \sum_{\theta=1}^{n_\text{c}}  W_R(\theta) W_L^*(\theta) p(\theta)= I_r,
\end{align}
where \(\theta\) denotes an index to specify the set of chosen rank-1 projectors, and \(p(\theta)\) is the probability for choosing \(\theta\).
(See \crefappendix{sec:sampling}{S4} for more details.)
Thus, the resulting tensor network with stochastic projectors defines an unbiased estimator for the partition function~\cite{ferrisUnbiasedMonteCarlo2015}.

However, there is a fatal flaw with the approach in \citeref{ferrisUnbiasedMonteCarlo2015}.
The number of projectors, \(N_\text{p}\), is in the same order as the number of tensors in the original network.
The variance of the contraction of a tensor network containing numerous random tensors explodes exponentially to \(N_\text{p}\), as it is essentially equivalent to a product of an extensive number of random numbers.
Such naive importance sampling only works for relatively small systems as the estimator's variance diverges exponentially with the system size.
(See \crefappendix{sec:product}{S1} for more details).
We must introduce a method that can control the variance, such as the sequential Monte Carlo~\cite{doucetSequentialMonteCarlo2001, arulampalamKalmanFilterParticle2004}, with resampling particles, or MCMC~\cite{newmanMonteCarloMethods1999,landauGuideMonteCarlo2014}.
In many cases, we are interested in the expected value of physical quantities, not the partition function itself, so the latter, MCMC, is the method of choice.

In the following, we denote the state of each stochastic projector as \(\theta_i\).
We aim to implement MCMC based on a simple Metropolis-Hasting (MH) scheme~\cite{metropolisEquationStateCalculations1953, hastingsMonteCarloSampling1970} with \(\{\theta_i\}\) as the state variables.
To do this, we need to introduce the following additional techniques.

\textit{Independent proposal distribution}:
In the original proposal by Ferris~\cite{ferrisUnbiasedMonteCarlo2015}, a set of rank-1 projectors is obtained based on the results of the previous short-scale renormalization step, and then importance sampling is performed.
In this method, when the short-scale projectors (e.g., red projectors in \cref{fig:trgp-8}) are updated, the definition of longer-scale projectors (blue and orange ones) changes, so it is difficult to implement MCMC.
In our method, on the other hand, we first execute the conventional deterministic tensor renormalization procedure (in the projector formulation) as an initialization process.
At the same time, we determine all the rank-1 projectors and their weights using the results of the SVD in that process.
During the MCMC process, we do not change the projector sets and the weights.
This initialization scheme makes the proposal distribution of the MCMC projectors independent for each projector, and the partition function can be written as
\begin{align}
  Z = \sum_{\{\theta_i\}} g(\theta_1,\theta_2,\ldots\theta_{N_\text{p}}) p(\theta_1) p(\theta_2) \cdots p(\theta_{N_\text{p}}).
  \label{eq:partition-function}
\end{align}
In addition, since SVD is only performed for initialization, and in the MCMC process, only the selection of rank-1 projectors and the contraction of the tensor network with embedded projectors is executed, it has the advantage of being very high-performance and executable on modern computer architectures.

\textit{Computational graph}:
The weight factor, \(g(\{\theta_i\})\) in \cref{eq:partition-function}, is the tensor network contraction with embedded projectors.
When the projector configuration changes, the contraction of the whole network must be evaluated again.
If we re-evaluate the entire tensor network every time, the computational complexity is \(O(d^5 N)\); that is, the cost of one MCMC sweep becomes \(O(d^5 N^2)\), which is very inefficient.
To reduce the order of this computational complexity, we introduce the concept of a computational graph.
The series of contractions can be represented as a tree structure.
The initial tensors and projectors correspond to the leaves of the tree structure.
When a projector, i.e., a pair of \(W_L(\theta)\) and \(W_R(\theta)\), is updated, only their ancestors need to be re-evaluated.
This way, in the case of TRG, the computational complexity of a single MCMC sweep is reduced to \(O(d^5 N \log N)\).

\begin{figure}[tbp]
  \noindent (a) \hfill \vspace*{-2em} \\
  \includegraphics[width=0.9\linewidth]{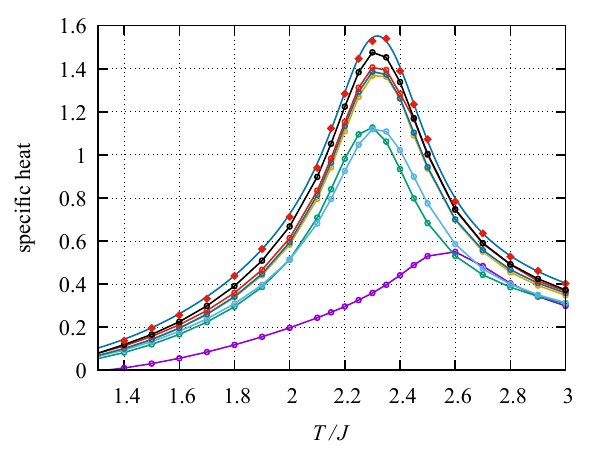} \\
  \noindent (b) \hfill \vspace*{-2em} \\
  \includegraphics[width=0.9\linewidth]{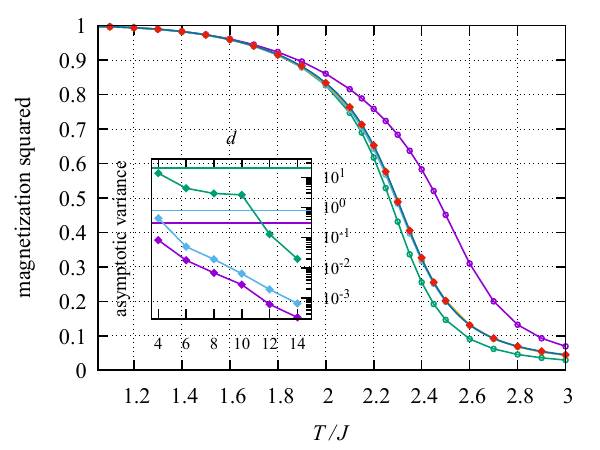} \\[-1em]
  \caption{%
    Temperature dependence of the specific heat (a) and the magnetization squared (b) of the Ising model on the \(N=16 \times 16\) square lattice.
    Dark blue curves represent the exact results of the transfer matrix calculation.
    Red diamonds represent the results of MCMC in the tensor network representation with \(d=6\).
    We also show results of the Levin-Nave TRG with impurity tensors with \(d=2\) (purple), 4 (green), 6 (cyan), 8 (orange), 10 (dark green)\(,\ldots,\) 16 (black). Inset in (b) represents the \(d\)-dependence of the asymptotic variance of the energy (purple), the specific heat (green), and the magnetization squared (cyan), together with those by the standard MH method (horizontal lines).
  }\label{fig:ising}
\end{figure}

\textit{Physical quantities}:
In the standard tensor network algorithms, finite difference or automatic differentiation of the free energy is used to evaluate the expectation value of the physical quantities, such as the internal energy and magnetization.
In our MCMC algorithm, we adopt the impurity tensor method~\cite{corbozVariationalOptimizationInfinite2016, moritaCalculationHigherorderMoments2019, moritaMultiimpurityMethodBondweighted2025} instead.
In the impurity tensor method, we prepare ``impurities,''  which are obtained by differentiating the initial tensors concerning the external variables, such as the temperature and the magnetic field, and then recursively evaluate the contraction of the tensor network that contains only one or two impurities in total, depending on the order of differentiation.
This method is easily implemented in projector-base tensor network renormalization and is compatible with our MCMC formalism.
One of the drawbacks of the impurity tensor method in ordinary tensor renormalization groups is that it does not consider projector derivatives, which introduces extra systematic errors.
However, in our MCMC method, the projectors become the identity operators after taking the random average and do not depend on external variables; we have no systematic error in the impurity tensor method.


To demonstrate the effectiveness of the proposed method, we present simulation results of the two-dimensional square lattice Ising model, which Hamiltonian is given by
\begin{align}
 H = -J\sum_{\langle i, j \rangle} \sigma_i \sigma_j - h \sum_i \sigma_i,
 \label{eq:ising}
\end{align}
where \(\sigma_i = \pm 1\) is the Ising spin variable at the \(i\)-th site, \(J\) is the coupling constant, and \(h\) is the magnetic field.

\begin{figure}[tbp]
  \centering
  \includegraphics[width=0.9\linewidth]{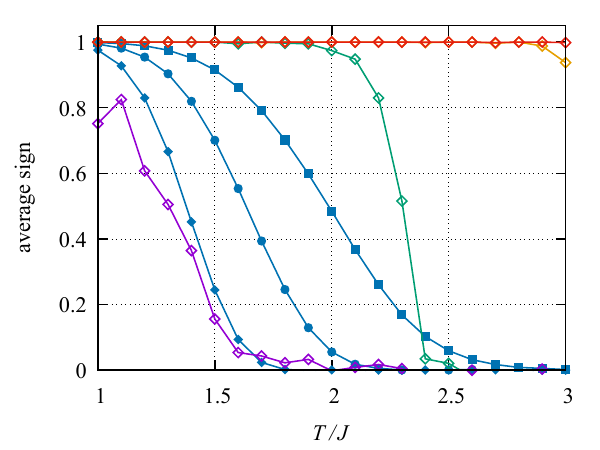} \\[-1em]
  \caption{%
    Temperature dependence of the average sign. Blue filled symbols represent the standard MCMC results for \(N=8 \times 8\) (squares), \(16 \times 16\) (circles), and \(32 \times 32\) (diamonds). Open diamonds represent those of MCMC in the tensor network representation for \(N=32 \times 32\). The bond dimension cutoff is \(d=2\) (purple), 3 (green), 4 (orange), and 6 (red).
  }\label{fig:negative}
\end{figure}

First, we present the results at zero magnetic field, \(h=0\).
In \cref{fig:ising}, we show the specific heat and the magnetization squared as functions of the temperature.
The system size is \(N = 16 \times 16\), and the bond dimension cutoff is set to \(d=6\).
The total MC steps are \(2^{14}\), of which the first \(2^{11}\) steps are discarded as burn-in time.
For comparison, we also show the exact results obtained by the transfer matrix method and the results by the Levin-Nave TRG with impurity tensors with \(d=2, 4, 6, 8, \ldots, 16\).
Note that \(d=256\) is required for TRG to obtain the exact result for \(N = 16 \times 16\).
The TRG results suffer from a systematic error due to the finite bond dimension cutoff and the impurity tensor method.
Furthermore, the convergence to the exact result is non-monotonic.
On the other hand, the MCMC results using \(d=6\) are fully consistent with the exact results, supporting that the systematic error is eliminated.

In the inset of \cref{fig:ising}, we show the \(d\)-dependence of the asymptotic variance, \(2\sigma_0^2\tau_\text{int}\) in \cref{eq:variance}, of the internal energy, specific heat, and magnetization squared at the critical temperature, \(T_c = 2J/\log(1+\sqrt{2})\)~\cite{kramersStatisticsTwoDimensionalFerromagnet1941}.
The asymptotic variance is evaluated using the binning analysis.
The variance is smaller than the conventional MH method already at \(d=4\) and decreases exponentially with the bond dimension cutoff \(d\).
Considering our proposed method's computational complexity is \(O(d^5 N \log N)\), the present results indicate that exponential acceleration has been achieved.

Next, we present the result at a pure imaginary magnetic field, \(h= i\pi/2\beta\), or a negative fugacity, \(z = e^{-2\beta h} = -1\).
This magnetic field is the intersection of the unit circle \(|z|=1\) of the Yang-Lee zeros~\cite{yangStatisticalTheoryEquations1952} and the negative real axis of \(z\).
At \(z=-1\), the Boltzmann weight, in the original spin-base representation, is given by \(W(\{\sigma_i\}) = (-1)^{m/2} e^{\beta \sum \sigma_i \sigma_j}\) with \(m\) being the total magnetization and the standard MCMC method suffers from the severe sign problem due to the factor \((-1)^{m/2}\).
In \cref{fig:negative}, we show the temperature dependence of the average sign.
Our proposed method, where the average sign is given by \(\langle \sgn g(\{\theta_i\}) \rangle\), also has negative signs at small \(d\).
However, in contrast to the standard MCMC method, the average sign is improved systematically and approaches unity rapidly as we increase \(d\).
Thus, the negative sign problem has also been prevented by paying a polynomial cost.


To summarize, we have introduced a novel MCMC framework in tensor network representations.
By combining projector formulation of the tensor network renormalization and MCMC sampling, we eliminate systematic error arising from finite bond dimension cutoff while maintaining the computational efficiency of tensor network techniques.
We demonstrated the effectiveness of our approach through simulations of the two-dimensional Ising model, showing exponential reductions in statistical error and a systematic resolution of the negative sign problem with polynomial computational cost. This framework is versatile and applicable to various sophisticated tensor network algorithms, such as HOTRG~\cite{xieCoarsegrainingRenormalizationHigherorder2012}, TEBD~\cite{vidalEfficientSimulationOneDimensional2004, verstraeteMatrixProductDensity2004}, ATRG~\cite{adachiAnisotropicTensorRenormalization2020}, BTRG~\cite{adachiBondweightedTensorRenormalization2022}, and CATN~\cite{panContractingArbitraryTensor2020}.

The proposed method facilitates accurate evaluation of partition functions, physical quantities, and tensor network optimization, making it a powerful tool for tackling complex physics problems.
Future work could explore extensions to higher-dimensional systems, applications in real-time unitary dynamics, and integration with advanced MCMC techniques, such as the extended ensemble methods~\cite{jankeMulticanonicalMonteCarlo1998, hukushimaExchangeMonteCarlo1996, diaconisAnalysisNonreversibleMarkov2000, turitsynIrreversibleMonteCarlo2011} and irreversible transition kernels~\cite{creutzOverrelaxationMonteCarlo1987, suwaMarkovChainMonte2010, bernardEventchainMonteCarlo2009, michelGeneralizedEventchainMonte2014}.
By offering both accuracy and computational efficiency, the proposed method represents a significant step forward in developing robust and scalable sampling techniques for computational science.

\begin{acknowledgments}
  The author thanks Hidemaro Suwa, Tsuyoshi Okubo, Xun Zhao, and Shimpei Goto for fruitful discussions and comments.
  This work was supported by JSPS KAKENHI Grant Numbers 20H01824 and 24K00543, the Center of Innovation for Sustainable Quantum AI (SQAI), JST Grant Number JPMJPF2221, and JST CREST Grant Number JPMJCR24I1.
\end{acknowledgments}

\appendix
\section{Product of Random Numbers}
\label{sec:product}

In this \AppendixName, we focus on the problem of sampling the product of $N$ independent random numbers.
Suppose that we have $N$ random variables, \(\{X_i\}_{i=1}^N\), that are independent and each of which is distributed according to probability distributions with mean \(\mu_i\) and variance \(\sigma_i^2\).
We are interested in estimating the expectation of the product of these random variables, \(\text{E}[X_1 X_2 \cdots X_N]\).
As \(\{X_i\}\) are independent of each other, the expectation of the product is given by the product of their expectations:
\begin{align}
 \text{E}[X_1 X_2 \cdots X_N] = \text{E}[X_1] \text{E}[X_2] \cdots \text{E}[X_N] = \prod_i \mu_i \equiv \Gamma.
\end{align}
The sample mean of the product of \(N\) random variables is an unbiased estimator of \(\Gamma\).
However, the relative statistical error diverges exponentially with \(N\):
\begin{align}
    \begin{split}
 \text{Var}[X_1 X_2 \cdots X_N] &= \text{E}[X_1^2 X_2^2 \cdots X_N^2] - \text{E}[X_1 X_2 \cdots X_N]^2 \\
        &= \text{E}[X_1^2] \text{E}[X_2^2] \cdots \text{E}[X_N^2] - \Gamma^2 \\
        &= \prod_i (\mu_i^2 + \sigma_i^2) - \Gamma^2 \\
        &\approx \Gamma^2 \prod_i (1 + \sigma_i^2/\mu_i^2) \qquad \text{for \(N \gg 1\)}.
    \end{split}
    \label{eqn:variance}
\end{align}

The divergence of the relative error can also be explained from a different viewpoint.
The multiplication of \(N\) random variables is interpreted as an \(N\)-step random walk on the logarithmic axis as
\begin{align}
    \log (X_1 X_2 \cdots X_N) = \sum_{i=1}^N \log X_i.
    \label{eqn:random-walk}
\end{align}
The linear increase of the variance of the random walk causes exponential divergence of the relative error in the estimate of \(\Gamma\).

\begin{figure}[tbp]
    \centering
    \includegraphics[width=0.9\linewidth]{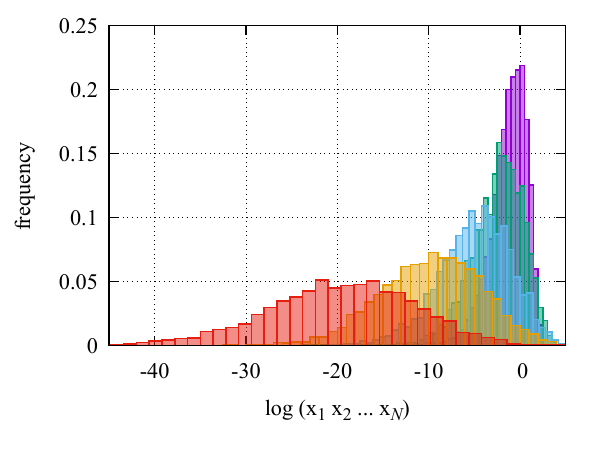}
    \caption{%
    Histogram of \(\log (x_1 x_2 \cdots x_N)\) for \(N=4\) (purple), 8 (green), 16 (cyan), 32 (orange), and 64 (red), where \(x_i\) are independently sampled from uniform distributed in \((0,2]\).
    The distributions exhibit biased random walk behavior with a peak at \((\log 2 - 1)N\) of width \(\sqrt{N}\).
 }
    \label{fig:hist}
\end{figure}

\begin{figure}[tbp]
    \centering
    \includegraphics[width=0.9\linewidth]{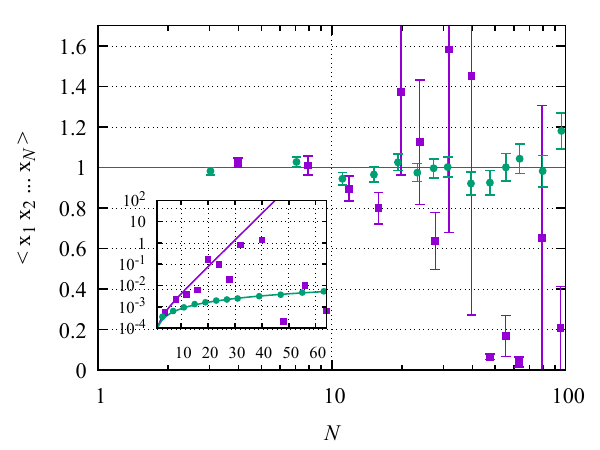}
    \caption{%
    \(N\)-dependence of the sample mean for the product of \(N\) random variables by the naive sequential sampling (purple) and those with multinominal resampling (green).
    The number of samples is \(M=2^{24}\) in both cases.
    Inset shows the \(N\)-dependence of the variance of the sample mean in the logarithmic scale.
    After introducing the resampling, the variance is suppressed by orders of magnitude.
 }
    \label{fig:scaling}
\end{figure}
  
Let's consider a simple example: \(\{X_i\}_{i=1}^N\) are independent and identity distributed (i.i.d.), uniformly in \((0,2]\):
\begin{align}
 P(X=x) = \begin{cases}
        1/2 & \text{for \(0 < x \le 2\)} \\
        0 & \text{otherwise}.
    \end{cases}
    \label{eq:prob_uniform}
\end{align}
The mean and the variance are \(\mu=1\) and \(\sigma^2=1/3\), respectively.
The probability density function of \(Y = \log X\) is given by
\begin{align}
 P(Y=y) = \begin{cases}
 e^y & \text{for \(y \le \log 2\)} \\
        0 & \text{otherwise},
    \end{cases}
    \label{eq:prob_log}
\end{align}
and the mean and the variance are \(\text{E}[Y] = \log 2 - 1\) and \(\text{Var}[Y] = 1\), respectively.

In \cref{fig:hist}, we show the histogram of samples, \(\{\log (x_1 x_2 \cdots x_N)\}\), for \(N=4,8,\ldots,64\).
The number of samples is \(M=2^{24}\).
As expected from \cref{eqn:random-walk}, a biased random walk behavior is observed; the peak position shifts as \((\log 2 - 1)N\), while the distribution width increases as \(\sqrt{N}\).
In other words, in the original linear scale, the typical value of \(x_1 x_2 \cdots x_N\) is exponentially small as \(\exp[(\log 2 - 1)N]\) and rare events with extremely large contributions dominate the mean \(\Gamma=1\).
In \cref{fig:scaling}, we plot the variance as a function of \(N\), which diverges exponentially as \((4/3)^N\) as predicted by \cref{eqn:variance}.
When \(N\) is large, the mean and the variance are greatly underestimated as the sample size is insufficient to pick up rare events.

Thus, even though as an estimator of the product of means, the sample mean of the product of the random variables is ``perfect,'' i.e., each sample is statistically independent and unbiased, it suffers from the exponential growth of statistical error when \(N\) is large.
This issue is the one that has surfaced in \citeref{ferrisUnbiasedMonteCarlo2015}.

The present naive perfect sampling is equivalent to the sequential Monte Carlo (SMC) method \textit{without} resampling.
In the SMC method~\cite{doucetSequentialMonteCarlo2001, arulampalamKalmanFilterParticle2004}, many particles, or walkers, are simulated simultaneously. 
To each particle, a weight is assigned.
These particles are updated according to the model.
However, a problem called ``weight degeneracy'' occurs, where a few particles gradually end up with extremely high weights while many others contribute almost nothing.
To prevent this phenomenon, we perform resampling during particle updates.
Resampling prunes particles with small weights and duplicates those with high weights to re-even out the overall distribution.
This technique prevents information from being concentrated in a few particles and increases the effective sample size.

Here, we adopt one of the simplest methods, multinominal resampling, for the present problem, where the weight of each particle is the product of the random numbers~\cite{gordonNovelApproachNonlinear1993}; first, we prepare \(M\) particles and initialize their weights to unity.
Then, each particle's weight is multiplied by a random number \(x\) at each step.
After the update process, we perform resampling so that the weights of all the particles become the same.
In the present problem, since there is no internal state in particles other than the weight, we replace the weight of all particles with their average.
This set of update and resampling is repeated \(N\) times to estimate the product's mean.
In \cref{fig:scaling}, we also plot the results after the introduction of resampling.
The suppression of the variance is remarkable.
The variance is now given by replacing \(\sigma^2\) in equation \cref{eqn:variance} by \(\sigma^2/M\), which is almost linear in \(N\):
\begin{align}
    \text{Var}[X_1 X_2 \cdots X_N] &= \Big(1 + \frac{\sigma^2}{M}\Big)^N - 1
    \approx \frac{\sigma^2 N}{M}
\end{align}
as long as \(M \gg N\).

Another established approach for controlling the variance of weights is the MCMC method~\cite{newmanMonteCarloMethods1999,landauGuideMonteCarlo2014}.
If we are interested in the weighted average of quantities, not the average weight, the MCMC method is more suitable.
In the main text, we adopt the MCMC method to estimate the average physical quantities in the canonical ensemble.

\section{Projector formulation of low-rank approximation}
\label{sec:svd}

Let's consider SVD, or rank-1 decomposition, of the product of two matrices, \(A \in \mathbb{C}^{r \times p}\) and \(B \in \mathbb{C}^{r \times q}\):
\begin{align}
  A^* B = U \Sigma V^* = \sum_{i=1}^s c_i \mu_i \nu_i^* ,
\end{align}
where \(\cdot^*\) denotes matrix conjugate transpose, \(s\) \(\le \min(p,q,r) \) is the rank of \(A^* B\), \(U \in \mathbb{C}^{p \times s}\) and \(V \in \mathbb{C}^{q \times s}\) are isometries (\(U^*U = V^* V = I_s\)), \(\Sigma \in \mathbb{R}^{s \times s}\) is a diagonal matrix whose diagonal elements are singular values \(\{c_i\}_{i=1}^s\), and \(\mu_i \in \mathbb{C}^p\) and \(\nu_i \in \mathbb{C}^q\) denote the \(i\)-th column vector of \(U\) and \(V\), respectively.
We assume that the singular values in \(\Sigma\) are sorted in descending order. 
For later convenience, we define a diagonal matrix \(\Lambda = \Sigma^{1/2}\).

Rank-\(d\) approximation gives a \(p \times q\) matrix of rank \(d\) (\(\le s\)) that approximates the input \(A^*B\).
By using the SVD introduced above, the optimal rank-\(d\) approximation that minimizes the Frobenius norm of the difference from the input is given by
\begin{align}
  A^* B \approx U \bar{\Lambda} (V\bar{\Lambda})^*.
  \label{eq:low-rank-approx}
\end{align}
Here, \(\bar{\Lambda} \in \mathbb{R}^{s \times d}\) is a matrix consisting of the first \(d\) columns of \(\Lambda\).
This procedure also defines the low-rank decomposition: \(A^* B \approx \bar{A}^* \bar{B}\) with \(\bar{A} = (U \bar{\Lambda})^* \in \mathbb{C}^{d \times p}\) and \(\bar{B} = (V \bar{\Lambda})^* \in \mathbb{C}^{d \times q}\).

The low-rank approximation can be represented in a different form.
We define \(P = W_R W_L^* \in \mathbb{C}^{r \times r}\) with
\begin{align}
  W_L &= A U (\bar{\Lambda}^+)^* , \\
  W_R &= B V (\bar{\Lambda}^+)^* ,
\end{align}
where \(\bar{\Lambda}^+ \in \mathbb{R}^{d \times s}\) is the Moore-Penrose inverse of \(\bar{\Lambda}\).
By noticing that \(\bar{\Lambda}^+ \Sigma = \bar{\Lambda}^*\),
one can confirm that \(P\) satisfies
\begin{align}
  P^2 &= (W_R W_L^*)^2 = W_R W_L^* = P, \\
  A^*PB &= A^* W_R W_L^* B = \bar{A}^* \bar{B}.
\end{align}
Thus, \(P\) is a projector and its insertion between \(A^*\) and \(B\) yields the rank-\(d\) approximation, \cref{eq:low-rank-approx}.

The projector \(P\) can be decomposed into the sum of rank-1 projectors as
\begin{align}
  P = \sum_{i=1}^d \eta_i \xi_i^* ,
\end{align}
where \(\xi_i\) and \(\eta_i\) are the \(i\)-th column vector of \(W_L\) and \(W_R\), respectively, i.e.,
\begin{align}
  \xi_i &= c_i^{-1/2} A \mu_i, \\
  \eta_i &= c_i^{-1/2} B \nu_i,
\end{align}
where \(\{\xi_i\}_{i=1}^d\) and \(\{\eta_i\}_{i=1}^d\) form a dual orthonormal basis set:
\begin{align}
  \xi_i^* \eta_j = \delta_{ij} \qquad \forall i,j \in [1,d],
\end{align}
or equivalently,
\begin{align}
  W_L^* W_R = I_d.
\end{align}

Note that by setting \(d = s\), one can recover the input matrix:
\begin{align}
  A^*PB = A^*B \qquad \text{when \(d = s\)}.
\end{align} 
However, \(P\) is not necessarily an identity matrix even in that case; if \(s < r\), the rank of \(P\) is smaller than \(r\), and it can not be an identity.

In such a case, we augment basis vectors to form a complete dual orthonormal basis set, \(\{\xi_i\}_{i=1}^r\) and \(\{\eta_i\}_{i=1}^r\); we add \(2(r-s)\) vectors to the basis set by using the Gramm-Schmidt orthonormalization.
After the augmentation, \(W_L\) and \(W_R\) become square, and thus 
\begin{align}
  P = W_R W_L^* = I_r
  \label{eqn:complete}
\end{align}
follows \(W_L^* W_R = I_r\).

In our MCMC method, we record these complete dual basis set \(\{(\xi_i, \eta_i)\}\) together with the corresponding singular values \(\{c_i\}\) during the initialization stage, where the standard tensor renormalization group is performed with optimal low-rank approximations.
The completeness property, \cref{eqn:complete}, is essential when introducing MC sampling of projectors, as the projectors will be inserted between matrices (or tensors), which are generated stochastically and different from those used in the initialization stage.

\section{Tensor Renormalization Group in Projector Formulation}
\label{sec:trg}

\begin{figure*}[tbp]
  \centering
  \includegraphics[width=0.9\linewidth]{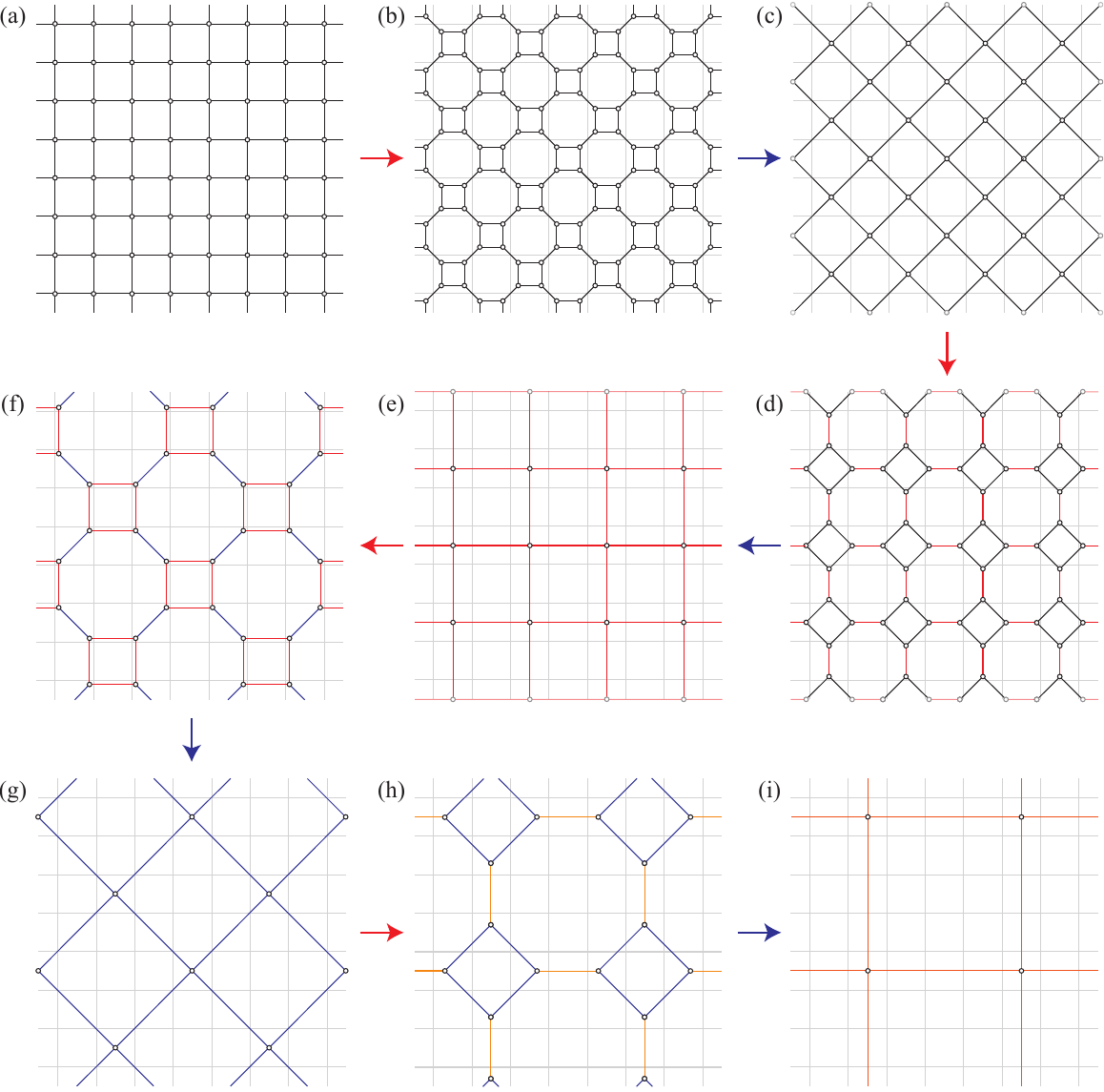}
  \caption{%
    Original Leven-Nave TRG procedure~\cite{levinTensorRenormalizationGroup2007}.
    Tensors in the original tensor network (a) are iteratively replaced by decomposition (red arrows) and contraction (blue arrows).
    We perform the low-rank approximation at the decomposition steps, keeping the bond dimension less than the cutoff \(d\) (red, blue, and orange bonds).
    After each set of decomposition and contraction, the number of tensors becomes half.
  }
  \label{fig:trg}
\end{figure*}

\begin{figure*}[tbp]
  \centering
  \includegraphics[width=0.9\linewidth]{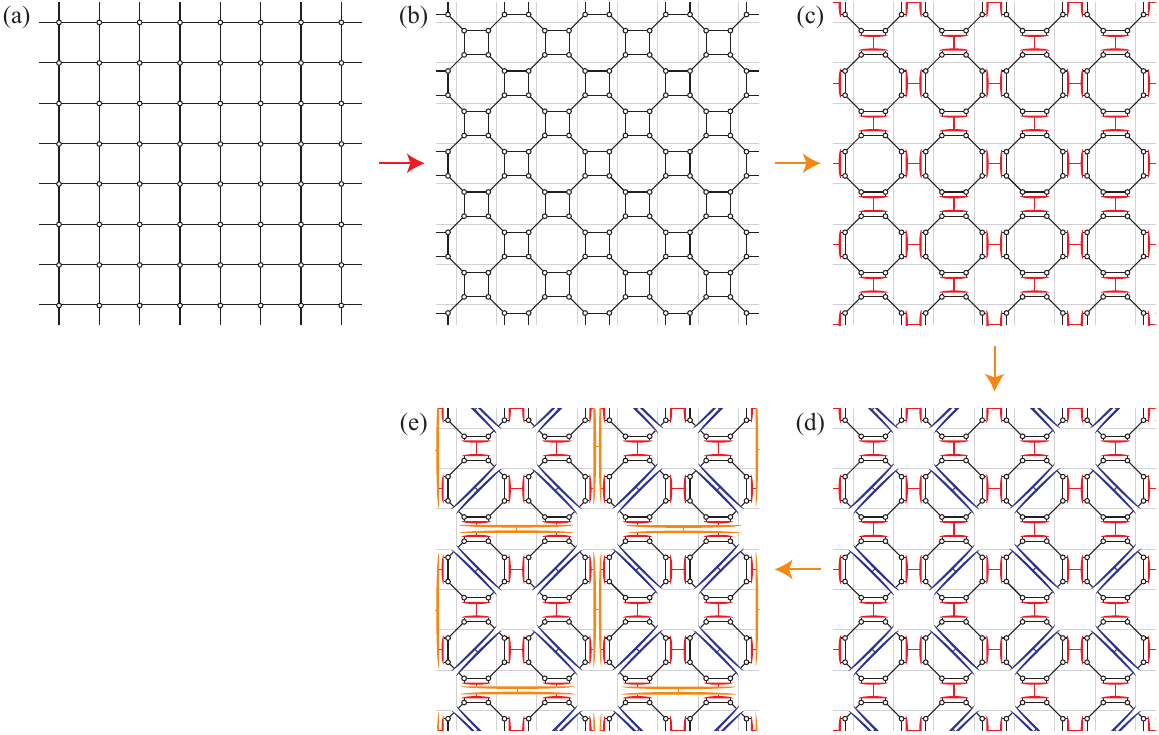}
  \caption{%
    TRG in projector formulation.
    The first step (b) is the same as the original TRG.
    Then, the next steps, (c) and (d) in \cref{fig:trg}, are replaced by the insertion of projectors (c), where pairs of three-leg tensors (red) are inserted between the tensors.
    We repeat the insertion of projectors ((d) and (e)).
    Note that the tree-leg tensors in (b) after the initial (exact) decomposition do not change during the renormalization steps.
  }
  \label{fig:trgp}
\end{figure*}

The partition function of the Ising model, \crefeq{eq:ising}{5}, can be represented by a tensor network.
First, we regard the Boltzmann weight of each bond term as a \(2 \times 2\) matrix:
\begin{align}
  W &= \begin{bmatrix}
    e^{\beta J} & e^{-\beta J} \\
    e^{-\beta J} & e^{\beta J}
  \end{bmatrix},
\end{align}
which is positive definite and can be decomposed as
\begin{align}
  W &= K^* K, \\
  \intertext{where}
  K &= \begin{bmatrix}
    \cosh^{1/2} \beta J & \cosh^{1/2} \beta J \\
    \sinh^{1/2} \beta J & -\sinh^{1/2} \beta J
  \end{bmatrix}.
\end{align}
By using \(K\), we define a four-leg site tensor as
\begin{align}
  T_{ijkl} &= \sum_m K_{im} K_{jm} K_{km} K_{l m} e^{\beta h \sigma_m}.
\end{align}
The partition function is then represented by a tensor network of \(T\) on the square lattice, as shown in \cref{fig:trg}(a).

At each iteration step of the Levin-Nave TRG~\cite{levinTensorRenormalizationGroup2007}, we decompose a four-leg tensor into two three-leg tensors [\cref{fig:trg}(b), (d), (f) (h)] and then contract four three-leg tensors to form a renormalized four-leg tensor [\cref{fig:trg}(c), (e), (g), (i)].
By repeating this procedure, all tensors are finally contracted to produce a scalar, the partition function.
The decomposition process is performed by SVD with the optimal low-rank truncation of the singular values (\crefappendix{sec:svd}).
In particular, the first decomposition step, \cref{fig:trg}(b), is performed explicitly as
\begin{align}
  T_{ijkl} &= \sum_m R_{ijm} R_{klm} ,\\
  \intertext{where}
  R_{ijm} &= K_{im} K_{jm} e^{\beta h \sigma_m / 2}.
\end{align}

The impurity tensors for calculating the internal energy and the specific heat 
(the magnetization and the susceptibility) is obtained by taking derivatives of \(R\) with respect to the inverse temperature \(\beta\) (the external field \(h\)).

The above tensor renormalization procedure can be rewritten in the projector formulation (\crefappendix{sec:svd}), where projector insertion replaces tensor contraction-and-decomposition.
The first step of the projector-base TRG is the same as the original TRG (\cref{fig:trgp}(b)).
Then, the next steps, \cref{fig:trg}(c) and (d), are replaced by \cref{fig:trgp}(c), where projectors, i.e., pairs of three-leg tensors (red), are inserted between the tensors.
This procedure is repeated as shown in \cref{fig:trgp}(d) and (e).
Note that in the original Levin-Nave TRG, tensors (denoted by black circles in \cref{fig:trg}) are transformed successively as the renormalization proceeds.
On the other hand, in the projector-base TRG, the tree-leg tensors in \cref{fig:trg}(b) remain unchanged until the final tensor network [\cref{fig:trg}(e)].

In the original TRG and its projector representation, we can use the translational symmetry of the system to reduce the \(N\)-dependence of the computational complexity from \(O(N)\) to \(O(\log N)\).
It should be noted, however, that when sampling projectors, the projectors at all locations must be treated as separate, or a systematic bias will be introduced~\cite{araiAllmodeRenormalizationTensor2023}.

As for the \(d\)-dependence of the computational complexity, the original TRG has \(O(d^6)\) complexity for the SVD of a four-leg tensor into two three-leg tensors.
However, by using the randomized SVD or the partial SVD, the complexity can be reduced to \(O(d^5)\)~\cite{halkoFindingStructureRandomness2011, moritaTensorRenormalizationGroup2018}.
In the initialization process of the proposed MCMC method, however, we need the full SVD to obtain all rank-1 projectors, whose complexity is still \(O(d^6)\).
In the later MCMC sampling stage, the complexity of the tensor network contraction is \(O(d^5)\).

\section{Sampling Projectors}
\label{sec:sampling}

At each step of MCMC, a projector, a set of rank-1 projectors, is proposed as a candidate according to predetermined trial probabilities; we choose a projector in \(r\) dimensions projecting onto a \(d\)-dimensional subspace.
A particular set of rank-1 projectors, \(\{k_1, k_2, \ldots, k_d\}\) (\(1 \le k_1 < k_2 < \cdots < k_d \le r\)), is chosen with a probability proportional to the product of weights of each rank-1 projector, \(\prod_{i=1}^d w_{k_i}\).
In this work, the weight of each rank-1 projector is defined by using the corresponding singular value \(c_i\) as
\begin{align}
  w_i = [\max(c_i, c_\text{max} \times 10^{-12})]^\omega,
\end{align}
where \(c_\text{max}\) is the largest singular value and the exponent \(\omega\) is a hyperparameter that controls the distribution of the weights.
For the numerical demonstration in the main text, we use \(\omega = 1\), but the optimal value \(\omega\) should be studied in more detail in the future.
There are \(\binom{r}{d}\) possible combinations to choose \(d\) out of \(r\).
In typical applications, \(r \approx d^2\).
For example, for \(d=32\), the number of combinations is more than \(10^{60}\). Thus, the naive tower sampling does not work at all.

For the present purpose, we can employ the dynamic programming~\cite{denardoDynamicProgrammingModels2003}.
We should notice that if \(m\) elements have already been chosen from \(\{1, \ldots, k-1\}\), the probability of choosing the \(k\)-th element is given by
\begin{equation}
  q^{(m)}_k = \frac{w_k R_{k,m+1}}{w_k R_{k,m+1} + R_{k,m}},
  \label{eqn:conditional-prob}
\end{equation}
where \(R_{k,m}\) (\(k=0, \ldots, r\); \(m=0, \ldots, d\)) is a sum of weight products of \(\binom{r-k}{d-m}\) combinations of choosing \((d-m)\) elements from \(\{k+1, \ldots, r\}\).
We can use this conditional probability to decide whether to adopt each element, \(k=1, \ldots, r\), one by one.
Also, the probability that a chosen set includes the \(k\)-th element \((k=1, \ldots, r)\) can be represented as
\begin{equation}
  q_k = \frac{s_k}{z_k}
  \label{eqn:marginal-prob}
\end{equation}
with
\begin{align}
  s_k &= \sum_{m=0}^{d-1} L_{k-1,m} w_k R_{k,m+1}, \label{eqn:s} \\
  z_k &= \sum_{m=0}^{d-1} L_{k-1,m} w_k R_{k,m+1} + \sum_{m=0}^{d} L_{k-1,m} R_{k,m}, \label{eqn:z}
\end{align}
where \(L_{k, m}\) (\(k=0, \ldots, r\); \(m=0, \ldots, d\)) is a sum of weights of \(\binom{k}{m}\) combinations of choosing \(m\) elements from \(1, \ldots, k\).
\(\{R_{k,m}\}\) and \(\{L_{k,m}\}\) are calculated by using the recursion relations:
\begin{align}
  R_{k,m} &=
  \begin{cases}
    \delta_{m,d} & \text{for \(k=r\)} \\
    w_{k+1} R_{k+1,m+1} + R_{k+1,m} & \text{otherwise,}
  \end{cases} \label{eqn:r-recursion}\\
  L_{k, m} &=
  \begin{cases}
    \delta_{m,0} & \text{for \(k=0\)} \\
    w_{k} L_{k-1,m-1} + L_{k-1,m} & \text{otherwise,}
  \end{cases} \label{eqn:l-recursion}
\end{align}
respectively, in advance at the cost of \(O(rd)\).
In applying \cref{eqn:r-recursion,eqn:l-recursion}, we should regard \(w_{0}\) and \(w_{d+1}\) as zero.
It is immediately clear from the above definitions that \(R_{k, d} = L_{k,0} = 1\) for \(k=1, \ldots, r\).
By using \cref{eqn:r-recursion,eqn:l-recursion}, \(z_k\) can be rewritten as
\begin{align}
  \begin{split}
  z_k &= \sum_{m=0}^{d-1} L_{k-1,m} (w_k R_{k,m+1} + R_{k,m}) + L_{k-1,d} R_{k,d} \\
  &= \sum_{m=0}^{d} L_{k-1,m} R_{k-1,m}.
  \end{split}
\end{align}
Note that \(z_k\), which is nothing but the partition function, does not depend on \(k\) and is equal to \(R_{0,0}\) or \(L_{r,d}\).
In the Appendix of \citeref{ferrisUnbiasedMonteCarlo2015}, an equivalent formulation has been presented in the tensor network language.

When applying the above expressions, additional care must be taken, as \(\{R_{k,m}\}\) and \(\{L_{k,m}\}\) are the sums of the products of many element weights, and overflow or underflow easily occurs if \(r\) becomes large.
One possible workaround is to use multi-precision floating-point arithmetic.
Alternatively, one can store the logarithm of the value and proceed with the log-sum-exponential trick.
In the following, we present a more natural implementation.
As we see below, our method is portable as it does not require multi-precision floating-point arithmetic and is faster as it does not require logarithmic and exponential operations.

First, we introduce new quantities \(\{S_{k,m}\}\) and \(\{T_{k,m}\}\), which are defined as the ratio of \(\{R_{k,m}\}\) and \(\{L_{k,m}\}\), respectively:
\begin{align}
  S_{k,m} &= 
  \begin{cases}
    R_{r,m} = \delta_{m,d} & \text{for \(k=r\)} \\
    1 & \text{for \(m=d\)} \\
    R_{k,m} / R_{k,m+1} & \text{for \(m<d\) and \(R_{k,m+1} > 0\)} \\
    0 & \text{otherwise,}
  \end{cases} \\
  T_{k,m} &=
  \begin{cases}
    L_{0,m} = \delta_{m,0} & \text{for \(k=0\)} \\
    1 & \text{for \(m=0\)} \\
    L_{k,m} / L_{k,m-1} & \text{if \(m>0\) and \(L_{k,m-1} > 0\)} \\
    0 & \text{otherwise.}
  \end{cases}
\end{align}
Note that \(\{S_{k,m}\}\) and \(\{T_{k,m}\}\) are quantities of the order of \(w_k\) unlike \(\{R_{k,m}\}\) and \(\{L_{k,m}\}\).
The following recursion relations manifest this fact:
\begin{align}
  S_{k,m} &= 
  \begin{cases}
    w_{k+1} + S_{k+1,d-1} & \text{for \(m=d-1\)} \\
    S_{k+1,m+1} \displaystyle \frac{w_{k+1} + S_{k+1,m}}{w_{k+1} + S_{k+1,m+1}} & \text{otherwise,}
  \end{cases} \\
  T_{k,m} &= 
  \begin{cases}
    w_{k} + T_{k-1,1} & \text{for \(m=1\)} \\
    T_{k-1,m-1} \displaystyle \frac{w_{k} + T_{k-1,m}}{w_{k} + T_{k-1,m-1}} & \text{otherwise.}
  \end{cases}
\end{align}
By using \(\{S_{k,m}\}\), the conditional probability \(q_k^{(m)}\) [\cref{eqn:conditional-prob}] is represented as
\begin{align}
  q_k^{(m)} &= 
  \begin{cases}
    \displaystyle \frac{w_{k}}{w_{k} + S_{k,m}} & \text{if \(w_k > 0\)} \\
    0 & \text{otherwise.}
  \end{cases}
\end{align}
Similarly, the marginal probability \(q_k\) [\cref{eqn:marginal-prob}] is represented using \(\{S_{k,m}\}\) and \(\{T_{k,m}\}\) as
\begin{align}
  \begin{split}
    q_k &= \frac{\displaystyle \sum_{m=0}^{d-1} \Big[\prod_{j=1}^m T_{k-1,j}\Big] w_k \Big[\prod_{j=m+1}^d S_{k,j}\Big]}{\displaystyle \Big[\prod_{j=0}^{d-1} S_{0,j} \Big]^{1/2} \Big[\prod_{j=1}^d T_{r,j} \Big]^{1/2}} \\
    &= w_k \sum_{m=0}^{d-1} \Big[\prod_{j=1}^m \frac{T_{k-1,j}}{S_{0,j-1}^{1/2}T_{r,j}^{1/2}}\Big] \Big[\prod_{j=m+1}^d \frac{S_{k,j}}{S_{0,j-1}^{1/2}T_{r,j}^{1/2}}\Big],
  \end{split}
\end{align}
which is evaluated at the cost of \(O(d)\) for each \(k\) and is free from overflow or underflow.

\bibliography{main,main-local}

\end{document}